# One-dimensional Cutting Stock Problem with Divisible Items

[1]Deniz Tanir, *[1]Onur Ugurlu, [2]Asli Guler, and [1]Urfat Nuriyev

[1]Ege University, Izmir, Turkey

tanirdeniz35@gmail.com, onurugurlu@mail.ege.edu.tr, urfat.nuriyev@ege.edu.tr

[2]Yasar University, Izmir, Turkey

asli.guler@yasar.edu.tr

**Abstract**. This paper considers the one-dimensional cutting stock problem with divisible items, which is a new problem in the cutting stock literature. The problem exists in steel industries. In the new problem, each item can be divided into smaller pieces, then they can be recombined again by welding. The objective is to minimize both the trim loss and the number of the welds. We present a mathematical model and a dynamic programming based heuristic for the problem. Furthermore, a software, which is based on the proposed heuristic algorithm, is developed to use in MKA company, and its performance is analyzed by solving real-life problems in the steel industry. The computational experiments show the efficiency of the proposed algorithm.

**Keywords:** One-dimensional cutting stock problem, heuristics, dynamic programming, nonlinear integer programming.

## 1. Introduction

The cutting-stock problem (CSP) has many applications in the production planning of many industries such as the metallurgy, plastics, paper, glass, furniture and textile. In general, cutting stock problems are based on cutting unlimited large pieces of length $c$ into a set of smaller items with demand $v_i$ and length $w_i$ while optimizing a certain objective function. The objective function can be minimizing waste, maximizing the profit, minimizing cost, minimizing the number of items used, etc [1].

The CSP has many practical applications in real-life problems and is easy to formulate. However, these problems are difficult to solve since they are in NP-hard [2]. Therefore, it is important to solve these problems efficiently so that the production expenses are lower as much as possible. In this paper, we consider the one-dimensional cutting stock problem with divisible items, which is a new problem in the cutting stock literature. The problem exists in the steel industry. In this problem, each item (demand) can be divided into smaller pieces, then they can be combined again by means of welding. The objective is to minimize both the trim loss and the number of the welds.

The remainder of the paper is organized as follows. In Section 2, we provide a formal definition of the classical one-dimensional cutting stock problem. We discuss the related studies in the literature in Section 3. We give the definition of the new problem in Section 4 and present the new mathematical model in Section 5. In Section 6, a dynamic programming-based heuristic for the problem is presented. In Section 7, the performance of the proposed algorithm is investigated on real life instances. The conclusion is provided in Section 8.

## 2. The Classical 1-D Cutting Stock Problem

The one-dimensional cutting stock (1D-CSP) problem is a well-known NP-hard problem [2] that occurs during manufacturing processes in many industries such as steel industry, clothing industry and aluminum industry; and has gained much attention from all over the world in recent years. The 1D-CSP is a linear integer programming problem with one decision variable for each possible pattern. If the number of order widths is small, then the number of patterns may be small enough that the problem can be solved using a standard algorithm. However, there may be an exponential number of patterns if the number of order widths is large. In these cases, an alternative solution approach is needed.

The classical 1D-CSP is the problem of cutting standard-sized pieces of stock material into pieces of specified sizes while minimizing the trim lost. In the standard definition of the problem the following data is used:

- $c$ stock lengths.
- $w_i$ order lengths of item $i$, $i=1,...,n$.
- $v_i$ demand number of item $i$.
- $x_{ij}$ number of items $i$ having been cut from stock $j$.
- $y_j$ indicates whether the stock $j$ is used in the cutting plan ($y_j=1$ if stock $j$ is used in the cutting plan), $j=1,...,m$.
- $t_j$ indicates the length of the leftover of stock $j$.

The objective is to minimize trim loss (wastage). Using the notation above, the classical 1D-CSP can be formulated as follows [3]:

$$\min \sum_{j=1}^{m} t_j$$

s.t.

$$\sum_{i=1}^{n} x_{ij} w_j + t_j = c \cdot y_j, \quad j=1,\ldots,m.$$

$$\sum_{i=1}^{n} x_{ij} = v_i, \quad j=1,\ldots,m.$$

$$x_{ij} \geq 0 \text{ integer}, \quad i=1,\ldots,n;\ j=1,\ldots,m.$$

$$t_j \geq 0 \text{ integer}, \quad j=1,\ldots,m.$$

$$y_j \in \{0,1\}.$$

In the model, the first constraint calculates the cutting waste of each stock in the cutting process, the second constraint guarantees that all demands for each item are met. And, the objective function minimizes the total trim loss.

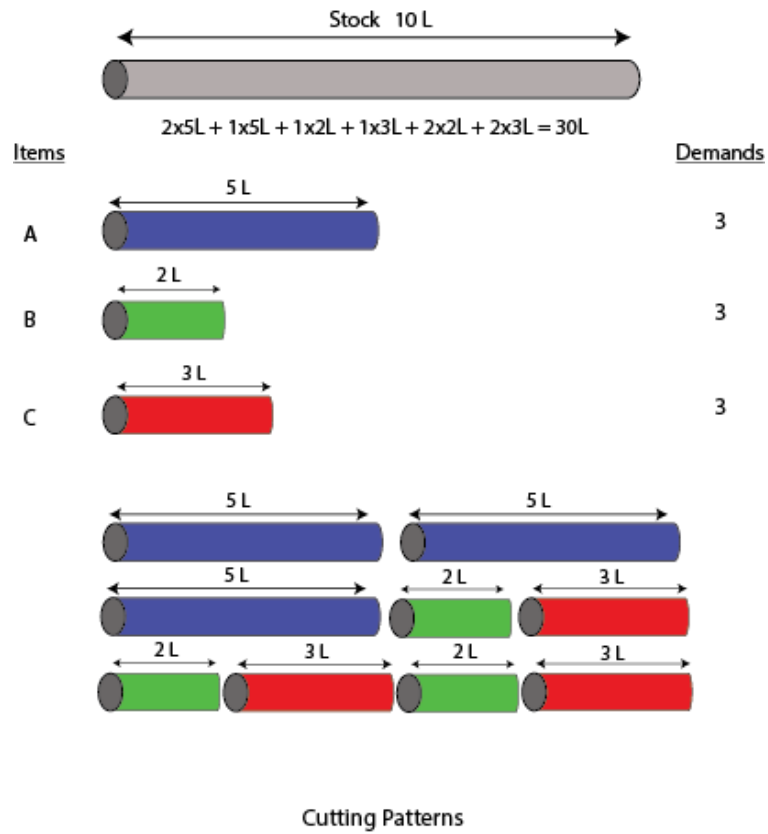

Fig. 1. 1D-CSP with one stock type.

## 3. Literature Review

The first known formulation of CSP was presented in 1939 by Kantorovich [4]. The most important advance in solving the 1D-CSP was the inventive work of Gilmore and Gomory [5,6], where they proposed delayed pattern generation method to solve the problem by using linear programming. Dyckhoff [7] classifies the solutions of CSP into two groups: pattern-oriented and item-oriented. Moreover, he classifies cutting problems by using four characteristics: dimensionality, kind of assignment, assortment of large objects, assortment of small objects.

Waescher and Gau [8] concluded that optimal integer solutions could be obtained in most cases. Gradisar et. al. [9] presented sequential heuristic procedure to optimize the 1D-CSP when all stock lengths are different. Furthermore, they proposed a hybrid approach which combines the pattern-oriented LP method and item-oriented sequential heuristic procedure [10]. Vance et al. [11], Vance [12], Vanderbeck [13,14] and Valerio de Carvalho [15] presented some attempts at combining column generation and branch-and-bound. They were able to obtain exact solutions for quite large instances of CSPs.

Scheithauer et al. [16] presented a cutting plane algorithm to solve the 1D-CSP exactly. Mukhacheva et. al. [17] proposed a modified branch-and bound method for 1D-CSP. Belov and Scheithauer [18] proposed an approach combining a cutting plane algorithm with column generation for the 1D-CSP with multiple stock lengths. Umetani et al. [19] considered that the number of different cutting patterns in the 1D-CSP is limited within a given bound. Then, they proposed an approach based on metaheuristics and incorporates an adaptive pattern generation method. Johnston and Sadinlija [20] generated a new model that resolves the non-linearity in the 1D-CSP, between pattern variables and pattern run-lengths by a novel use of 0-1 variables. Belov and Scheithauer [21] developed a branch-and-cut-and-price algorithm for one-dimensional stock cutting and two-dimensional two-stage cutting problems. They investigated a combination of the LP relaxation at each branch-and-price node is strengthened by Chvatal-Gomory and Gomory mixed-integer cuts.

In recent years, there have also been several efforts to solve this problem by using different approaches. Dikili et. al. [22] presented a successive elimination method to solve 1D-CSP, in ship production, directly by using the cutting patterns obtained by the analytical methods at the mathematical modeling stage. Reinertsen and Vossen [23] considered the CSP when orders have due dates and proposed new optimization models and solution procedures to solve this problem. Cherri et. al. [24] studied the cutting stock problem with usable leftovers and modified well-known heuristics in the literature to solve this problem. Furthermore, Cherri et. al. [25] assumed that the available retails in stock have priority-in-use during the cutting process and developed their heuristics considering these priorities. Berberler and Nuriyev [26] considered the subset-sum problem as a sub-problem to solve the 1D-CSP and proposed a dynamic programming-based heuristic. Mobasher and Ekici [27] studied a more general case of the classical CSP, called cutting stock problem with setup cost in which the objective is to minimize the total production cost including material and setup costs. They developed a mixed integer linear model and proposed an algorithm for a special case of the problem.

## 4. Definition of the New Problem

In this study, we consider a large manufacturer of steel profiles. In the steel industry, smaller-sized items are usually supplied from a standard length of 6000 mm or 12000 mm. The goal of companies is to determine the optimal cutting plan for minimizing the trim loss. Since minimizing trim-loss is a vital matter in the steel industry, a company has tried to divide items into two smaller pieces and recombined them again. When they encounter a large leftover from a cutting pattern, they try to use the leftover as a small piece of an item (demand). Then, they try to cut the residual part of the item in another cutting pattern, and they recombine the pieces again by the help of welding. To be able to perform welding, these pieces must be greater than a pre-set upper bound. (In our case, the bound for welding is 1000 mm). Moreover, only one piece can be used for a part of an undivided item in a single cutting pattern, but the residual parts of the divided items can be used as required. (Note that this constraint is mainly valid for application purposes). However, the welding is an extra cost for the company. Thus, unlike to classical CSP, our problem has two objectives: The primary objective is to minimize the trim loss, and the secondary objective is to minimize the number of welds.

In the steel cutting, the leftovers obtained after cutting objects in stock can be stored for future use, if they are longer than a pre-set upper bound (In our case, the bound is 1000 mm). Thus, the company has tried to seek a solution to the cutting stock problem that will not only decrease trim loss in a period but also improve the total results over the whole time-span. In section 6, a heuristic algorithm which takes all constraints of companies into consideration is described.

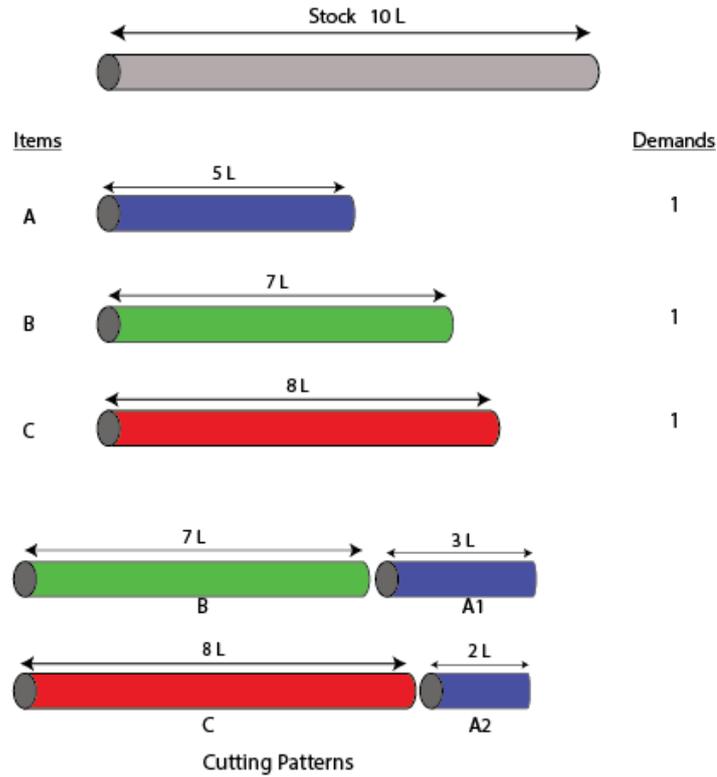

Fig. 2. 1D-CSP with divisible items.

The problem of optimal steel manufacturing in the company can be modeled as a 1D-CSP using following nonlinear integer programming.

## 5. The Formulation of the New Problem

In this section, a nonlinear integer programming formulation is presented for the described version of 1D-CSP. In the proposed new problem, we are given an unlimited number of stock (let $m$ be the number of stock) of length $c$ and $n$ different types of items.

To quantify the upper bounds for usable leftovers and welding, we use the parameters $\beta$ and $\theta$, defined by the user (decision maker).

- $\beta$ : upper bound for trim loss.
- $\theta$ : upper bound for welding.

$\beta$ and $\theta$ can be set according to the companies' requirement. The objective is to minimize both total trim loss and the number of the welds. The problem may be formulated as follows:

- $z_{ij}^k$ indicates whether $k^{th}$ piece of item $i$ is divided in stock $j$ ( $z_{ij}^k = 1$, if the $k^{th}$ piece of item $i$ is divided in stock $j$).
- $\alpha_{ij}^k$ indicates the length of the residual part of the $k^{th}$ divided piece of item $i$ in stock $j$.
- $b_{ij}^k$ indicates whether the residual part of the $k^{th}$ divided piece of item $i$ is used in stock length $j$ ( $b_{ij}^k = 1$, if the residual part of the $k^{th}$ divided piece of item $i$ is used in stock length $j$).
- $u_j$ indicates whether the leftover stock $j$ is counted as trim loss ($u_j = 0$, if the length of the leftover of stock $j$ is greater than $\beta$ and it is not counted as trim loss).

(1) $\min \sum_{i=1}^{m} t_j u_j$

(2) $\min \sum_{i=1}^{n} \sum_{k=1}^{v_i} \sum_{j=1}^{m} z_{ij}^k$

s.t.

(3) $\sum_{i=1}^{n} x_{ij} w_i + \sum_{i=1}^{n} \sum_{k=1}^{v_i} z_{ij}^k (w_i - \alpha_{ij}^k) + \sum_{i=1}^{n} \sum_{k=1}^{v_i} \sum_{l=1, l \neq j}^{m} b_{ij}^k \alpha_{il}^k + t_j = c y_j, \quad j = 1, ..., m.$

(4) $\sum_{j=1}^{m} x_{ij} + \sum_{k=1}^{v_i} \sum_{j=1}^{m} z_{ij}^k = v_i, \quad i = 1, ..., n.$

(5) $\sum_{k=1}^{v_i} \sum_{j=1}^{m} z_{ij}^k = \sum_{k=1}^{v_i} \sum_{j=1}^{m} b_{ij}^k, \quad i = 1, ..., n.$

(6) $\sum_{i=1}^{n} \sum_{k=1}^{v_i} z_{ij}^k \leq 1, \quad j = 1, ..., m.$

(7) $\alpha_{ij}^k \geq \theta, \quad i = 1, ..., n, j = 1, ..., m, k = 1, ..., v_i.$

(8) $w_i - \alpha_{ij}^k \geq \theta, \quad i = 1, ..., n, j = 1, ..., m, k = 1, ..., v_i.$

(9) $u_j = \begin{cases} 0 & \text{if } t_j \geq U \\ 1 & \text{otherwise.} \end{cases}$

(10) $\alpha_{ij}^k \geq 0$ integer, $i = 1, ..., n; k = 1, ..., v_i; j = 1, ..., m.$

$x_{ij} \geq 0$ integer, $i = 1, ..., n; j = 1, ..., m.$

$t_j \geq 0$ integer, $j = 1, ..., m.$

$y_j \in \{0,1\}, z_{ij}^k \in \{0,1\}, b_{ij}^k \in \{0,1\}, u_j \in \{0,1\}.$

The objective (1) minimizes the sum of trim loss and the objective (2) minimizes the number of welds. The optimal cutting pattern for each of the stock is defined by decision variables $b_{ij}^k$, $z_{ij}^k$, $\alpha_{ij}^k$, $x_{ij}$ in the constraint (3). By the constraint (4), all demands for each item are met. The constraint (5) guarantees that residual part of divided items are used. The constraint (6) controls that only one undivided item can be divided and used in a single cutting pattern. The constraint (7) and (8) control that the lengths of the pieces of a divided item are greater than the pre-set upper bound $\theta$. Finally, the constraint (9) determines whether the leftover of stocks can be used in the future.

## 6. The Proposed Algorithm

It is quite obvious that the new problem is NP-Hard in the strong sense like the classical 1D-CSP [2]. Since we are dealing with real life instances, solving those large instances with the proposed nonlinear integer programming formulation can take long running times. Therefore, we proposed a fast heuristic to obtain high-quality solutions in reasonable times. These heuristic procedures were obtained modifying the heuristics (BBP) proposed by Berberler and Nuriyev [26].

The BBP algorithm considers the subset problem as a sub-problem of 1D-CSP. At each step, the algorithm tries to find a cutting pattern via dynamic programming. However, unlike the standard dynamic programming, the algorithm uses only one one-dimensional array $A$ instead of two arrays with dimension $n \times c$. To construct $A$, the following function $F_k(s)$ is used;

$$F_k(s) = \max_x \{\sum_{i=1}^{k} w_i x_i \mid \sum_{i=1}^{k} w_i x_i \leq s,\ 0 \leq x_i \leq v_i,\ x_i \in N,\ i = \overline{1,k}\}$$

where $(s = \overline{1,c},\ k = \overline{1,n})$

Initial values of this function ($s = \overline{1,c},\ k = 1$) are determined as follows;

$$F_1(s) = \begin{cases} (1, s/w_1), & \text{if } s \leq v_1 w_1 \text{ and } s/w_1 = \lfloor s/w_1 \rfloor \\ (0,0), & \text{otherwise} \end{cases}$$

The recursive formula for $k>1$ occurs as below, ($k = \overline{2,n}, \ s = \overline{1,c}$)

$$F_k(s) = \begin{cases} F_{k-1}(s), & \text{if } F_{k-1}(s) \neq 0 \text{ or } s < w_k \text{ or } s > \sum_{j=1}^{k} w_j v_j \\ (k, p+1), & \text{if } F_k(s-w_k) = (k, p) \text{ and } p < v_k \\ (k, 1), & \text{if } s = w_k \text{ or } F_k(s-w_k) \neq 0 \end{cases}$$

After dynamic programming performed, if the last cell of the array is filled, this means that the stock is filled. Otherwise, the full cell with maximum number yields an optimal solution. This procedure continues until all demands are completed. For detailed information about the algorithm, the reader is referred to Berberler and Nuriyev [26]. To modify BBP according to our new problem, we use an another array $B$ with the same size with array $A$, and both of these arrays are constructed simultaneously. While array $A$ is being constructed according to above function $F_k(s)$, array $B$ try to improve solutions, which are found in array $A$, by finding a proper smaller piece of an item at each step. If a smaller piece of an item fulfills the stock when it added to the solution, then it is marked in array $B$.

Since our problem has two main objectives, we use following parameters $\gamma$ and $\delta$ to decide which cutting pattern will be used.

- $\gamma$ : the cost of 1 mm trim loss.
- $\delta$ : the cost of one welding operation.

After $A$ and $B$ are constructed, the cutting pattern which has the minimum cost for the company, is selected from the arrays. The residual parts of the divided items, are considered as a new single item in next steps and they are used in the construction of $A$. This procedure continues until all demands are met. It is easy to see that for small $\delta$ values, the trim loss converges to zero. However, for bigger $\delta$ values, the algorithm will find a solution which has slightly welding operation but a lot more trim loss. Parameters can be set in many different ways in line with the goal of the company. The advantage of the proposed approach is to determine an optimal cutting pattern at each step. Also, the best cutting pattern which includes a divided item is found at each step. By minimizing both trim loss and the number of welds, the proposed method can capture the ideal solution of the analytical methods. A computer program, which is based on the proposed algorithm, was developed to use in MKA company [28]. MKA is founded at 2005 May and the aim of the company is to be specialist on engineering software. Now, MKA uses our algorithm to create cutting plans for steel manufactories.

## 7. Computational Experiments

To the best of our knowledge, this version of the 1D-CSP has been studied for the first time. Thus, we only reported the results of the algorithm for six difficult real-life optimization problems that are taken from MKA. These problems can be downloaded via [29]. The length of the stock $c$ is 12000 mm for each of the six problems. Computational experiments run through Intel Core i7 4700HQ 2.4 GHz CPU with 32GB of RAM running Windows 8.1 64-bit Edition. The parameters $\gamma$ and $\delta$ are set to 1 and 500, respectively. Table 1 shows the results of the proposed algorithm on the real life instances.

Table 1. Results of the proposed algorithm on the real life instances.

|    | n  | number of total demands | sum of all demands length | number of the used stocks | total trim loss | percentage of trim loss | number of the welds | usable leftovers | time in sec. |
|----|----|----|----|----|----|----|----|----|----|
| p1 | 8  | 41  | 128.926   | 21  | 501  | 0,228 | 11 | 7791*1  | 2  |
| p2 | 18 | 62  | 275.383   | 24  | 958  | 0,347 | 8  | 1008*1  | 7  |
|    |    |     |           |     |      |       |    | 10472*1 |    |
| p3 | 20 | 64  | 433.250   | 37  | 2466 | 0,569 | 18 | 2574*1  | 17 |
|    |    |     |           |     |      |       |    | 2897*1  |    |
| p4 | 23 | 316 | 1.247.293 | 107 | 3111 | 0,657 | 12 | 1350*1  | 14 |
|    |    |     |           |     |      |       |    | 4950*1  |    |
| p5 | 28 | 82  | 373.760   | 32  | 2457 | 0,657 | 14 | 7777*11 | 14 |
| p6 | 41 | 154 | 675.222   | 57  | 2836 | 0,42  | 32 | 5928*1  | 72 |

The experiments show that the proposed algorithm can find cutting plans with trim loss percentage <1 with an acceptable number of welds.

## 8. Conclusions

In this paper, we have presented a new version of the one-dimensional cutting stock problem, called Cutting Stock Problem with divisible items, in which our goal is to minimize both total trim loss and the number of welds in the steel industry. We have first developed a nonlinear integer formulation for this NP-hard problem. Also, a dynamic programming based heuristic has been designed to fulfill the companies' need. Moreover, a computer program which is based on the proposed algorithm has been developed to use in the MKA company. The computational experiments have shown that the algorithm performs well on real life instances.

**Acknowledgments**

The first and second authors gratefully acknowledge the support of TUBITAK (The Scientific and Technological Research Council of Turkey) 2211 program.

**References**

1. Kellerer, H., Pferschy, U., Pisinger, D.: Knapsack Problems. Springer-Verlag Berlin Heidelberg (2004).
2. Garey, M. R., Johnson, D. S: Computers and Intractability: A Guide to the Theory of NP-Completeness. WH Freemann, New York (1979).
3. Jahromi, M. H., Tavakkoli-Moghaddam, R., Makui, A., Shamsi, A.: Solving an one dimensional cutting stock problem by simulated annealing and tabu search. Journal of Industrial Engineering International. 8.1, 1-8, (2012).
4. Kantorovich, Leonid V. : Mathematical methods of organizing and planning production. Management Science 6.4, 366-422, (1960).
5. Gilmore, P.C., Gomory, R.E.: A linear programming approach to the cutting-stock problem. Operations Research. 9, 849-859, (1961).
6. Gilmore, P.C., Gomory, R.E.: A linear programming approach to the cutting-stock problem Part II. Operations Research. 11, 863-888, (1963).
7. Dyckhoff, H.: A typology of cutting and packing problems. European Journal of Operational Research. 44, 145-159, (1990).
8. Waescher, G., Gau, T.: Heuristics for the Integer one-dimensional Cutting Stock Problem. A Computational Study. OR Spektrum 18. 3, 131-144, (1996).
9. Gradiar, M., Kljaji M., Resinovi, G., Jesenko J.: A sequential heuristic procedure for one-dimensional cutting. European Journal of Operational Research. 114. 3, 557-568, (1999).
10. Gradisar M., Resinovic G., Kljajic, M.: A hybrid approach for optimization of one-dimensional cutting. European Journal of Operational Research. 119. 3, 719-728, (1999).
11. Vance, P., Barnhart, C., Johnson, E.L., Nemhauser, G.L.: Solving binary cutting stock problems by column generation and branch-and-bound. Computational Optimization and Applications. 3. 111-130, (1994).
12. Vance, P.: Branch-and-price algorithms for the one-dimensional cutting stock problem. Computational Optimization and Applications. 9, 211-228, (1998).
13. Vanderbeck, F.: Computational study of a column generation algorithm for bin packing and cutting stock problems. Mathematical Programming A. 86, 565-594, (1999).
14. Vanderbeck, F.: On DantzigWolfe decomposition in integer programming and ways to perform branching in a branch-and-price algorithm. Operations Research. 48, 111-128, (2000).

15. Valerio de Carvalho, J.M.: Exact solution of bin-packing problems using column generation and branch-and-bound. Annals of Operation Research. 86, 629-659, (1999).

16. Scheithauer, G., Terno, J., Muller, A., Belov, G.: Solving one-dimensional cutting stock problems exactly with a cutting plane algorithm. Journal of the Operational Research Society. 52, 1390-1401, (2001).

17. Mukhacheva, E. A., Belov, G,, Kartak, V., Mukhacheva, A. S.: One-dimensional cutting stock problem: Numerical experiments with the sequential value correction method and a modified branch-and-bound method. Pesquisa Operacional. 2000.2, 153-168, (2001).

18. Belov, G., Scheithauer, G.: A cutting plane algorithm for the one-dimensional cutting stock problem with multiple stock lengths. European Journal of Operational Research. 141, 274-294, (2002).

19. Umetani, S., Yagiura, M., Ibaraki, T.: One-dimensional cutting stock problem to minimize the number of different patterns. European Journal of Operational Research. 146, 388-402, (2003).

20. Johnston, R.E., Sadinlija, E.: A new model for complete solutions to one-dimensional stock problems. European Journal of Operational Research. 153, 176-183, (2004).

21. Belov, G., Scheithauer, G.: A branch-and-cut-and-price algorithm for one-dimensional stock cutting and two-dimensional two-stage cutting. European Journal of Operational Research. 171, 85-106, (2006).

22. Dikili A.C., Sarz E., Pek N. A.: A successive elimination method for one-dimensional stock cutting problems in ship production. Ocean engineering. 34.13, 1841-1849, (2007).

23. Reinertsen, H., Vossen Thomas W.M.: The one-dimensional cutting stock problem with due dates. European Journal of Operational Research. 201.3, 701-711, (2010).

24. Cherri, A.C., Arenales, M.N., Yanasse, H.H.: The one-dimensional cutting stock problems with usable leftover: a heuristic approach. European Journal of Operational Research. 196.3, 897-908, (2009).

25. Cherri, A.C., Arenales, M.N., Yanasse, H.H.: The usable leftover one-dimensional cutting stock problem-a priority-in-use heuristic. Intl. Trans. in Op. Res. 20, 189-199, (2013).

26. Berberler, M. E., Nuriyev, U. G.: A New Heuristic Algorithm for the One-Dimensional Cutting Stock Problem. Appl. Compuy. Math. 9.1, 19-30, (2010).

27. Mobasher A., Ekici A.: Solution approaches for the cutting stock problem with setup cost. Computers and Operations Research. 40, 225-235, (2013).

28. https://www.mkasteel.com/

29. http://fen.ege.edu.tr/~urfat/CSP/index.html